\documentclass[journal]{IEEEtran}

\usepackage{cite}
\usepackage{amsmath}
\usepackage{url}
\usepackage[utf8]{inputenc}
\usepackage{mwe}
\usepackage{subfig}
\usepackage{graphicx}
\usepackage{siunitx}
\usepackage[colorlinks,citecolor=blue]{hyperref}

\graphicspath{{figures/}}

\newcommand\varpm{\mathbin{\vcenter{\hbox{%
  \oalign{\hfil$\scriptstyle+$\hfil\cr
          \noalign{\kern-.3ex}
          $\scriptscriptstyle({-})$\cr}%
}}}}

\begin{document}

\title{Center-of-Mass Corrections in Associated Particle Imaging}

\author{Caroline~Egan,
        Ariel~Amsellem,
        Daniel~Klyde,
        Bernhard~Ludewigt,
        and~Arun~Persaud

\thanks{
All authors are with the Accelerator Technology \& Applied Physics Division of Lawrence Berkeley National Laboratory, Berkeley, CA 94720 USA;
\noindent e-mail: apersaud@lbl.gov}
}

%

\maketitle

\begin{abstract}
Associated Particle Imaging (API) utilizes the inelastic scattering of neutrons produced in deuterium-tritium fusion reactions to obtain 3-D isotopic distributions within an object. The locations of the inelastic scattering centers are calculated by measuring the arrival time and position of the associated alpha particle produced in the fusion reactions, and the arrival time of the prompt gamma created in the neutron scattering event. While the neutron and its associated particle move in opposite directions in the center-of-mass (COM) system, in the laboratory system the angle is slightly less than \qty{180}{\degree}, and the COM movement must be taken into account in the reconstruction of the scattering location. Furthermore, the fusion reactions are produced by ions of different momenta, and thus the COM velocity varies, resulting in an uncertainty in the reconstructed positions. In this article, we analyze the COM corrections to this reconstruction by simulating the energy loss of beam ions in the target material and identifying sources of uncertainty in these corrections. We show that an average COM velocity calculated using the ion beam direction and energy can be used in the reconstruction and discuss errors as a function of ion beam energy, composition, and alpha detection location. When accounting for the COM effect, the mean of the reconstructed locations can be considered a correctable systematic error leading to a shift/tilt in the reconstruction. However, the distribution of reconstructed locations also have a spread that will introduce an error in the reconstruction that cannot be corrected. In this article, we will use the known stopping powers of ions in materials and reaction cross sections to examine the reconstruction uncertainties. We also discuss the impact of this effect on our API system.

\end{abstract}

\section{Introduction}
\IEEEPARstart{I}NELASTIC Neutron Scattering (INS), in which a neutron scatters off a target nucleus and leaves the nucleus in an excited state, provides a unique way to analyze materials. When the nucleus relaxes back into its ground state, a gamma ray (or several) with a specific energy is produced which can be used to identify the target isotope. Associated Particle Imaging (API) is a specialized technique based on INS that allows for the 3-D reconstruction of the position of the neutron scattering center.

In an API system, neutrons and alpha particles are created in a deuterium-tritium (DT) fusion reaction (DD-API is also possible\cite{vainionpaa2013high}, but we focus on DT in this article). In the DT reaction, the energies of the neutron and its associated alpha particle, as well as their relative angle of \qty{180}{\degree} in the center-of-mass (COM) frame, are fixed. By placing a position-sensitive detector into the path of the alpha particle and by also measuring the detection times for the alpha particle and gamma ray, the INS location can be calculated. API therefore allows the position-resolved measurement of elemental distributions in a target region. A unique advantage of API systems is that due to the position measurement, background gamma rays from other areas can be removed. This leads to a highly improved signal-to-noise ratio for the target region. Current applications of API include the detection of explosives\cite{carasco2008field}, illicit drugs\cite{fontana2017detection}, special nuclear material\cite{HAUSLADEN2007387}, and diamonds\cite{alexakhin2015detection}. We are interested in applying API techniques in an agricultural context. Specifically, we see API as an exciting opportunity to provide accurate measurements of carbon distributions in the top \qtyrange{0}{30}{\cm} of soil and in the rapid measurement of the carbon content of soil cores. INS has previously been used to measure the first \qty{8}{\cm} of soil\cite{kavetskiy2019application}, but API allows us to extend measurements to deeper layers and therefore make more accurate estimates of the total carbon in acre-sized fields. We are developing an API-based instrument \cite{Unzueta_Mauricio2021-jq} for quantifying and monitoring carbon sequestration in soil \cite{Sanderman2017-rt}. Apart from carbon, API also measures distributions of other elements such as iron, aluminum, silicon and oxygen, which can provide helpful information about the soil. While we investigated the COM correction in the context of our API system primarily aimed at carbon-in-soil measurements, the study presented in this article is likely more important for other API applications that require high spatial resolution.

Investigations into COM corrections for API systems have  been conducted previously\cite{grogan2010development, cates2013investigation}, mainly to investigate the alpha-neutron angular distribution in a 1-D geometry considering the alpha detector and neutron generator. Here we provide further analysis into the errors introduced by the COM effect and its dependency on beam parameters, system geometry, 2-D alpha detection, and energy loss within the titanium target. We quantify these errors and identify ways in which to reduce them and examine the impact that the COM distribution has on our current API system.

In the following section, we explain the physics involved in API, describe the reconstruction algorithm, and lay out the methods used in our analysis. In the final sections, we present our results and conclusions.

\section{Methods}

An API system consists of an alpha detector, gamma detectors (in our case a 3" cylindrical LaBr$_3$ and a 5" cylindrical NaI detector), and a neutron generator (comprised of an ion source, an acceleration gap, and a neutron-production target). In our experiments, we used an API neutron generator built by Adelphi Technology\cite{adelphi}, model DT108API (with a tilted alpha detector), and therefore the calculations presented in this article are based on this design. 

Ions are generated from a microwave-driven plasma and electrostatically accelerated towards a the neutron production target which consists of a thin titanium layer on a copper backing. To minimize the amount of material the alpha particle has to travel through (and therefore avoiding potential alpha particle scattering), the neutron production target and the alpha detector are angled with respect to the incoming ion beam. As we will see later, tilting the alpha detector also helps to reduce COM effects. Once the target is beam-loaded with accelerated deuterium and tritium ions, subsequent ions collide with deuterium/tritium nuclei embedded in the titanium layer resulting in nuclear fusion reactions of the form
\begin{equation}
    D + T = n\,(\qty{14.1}{\MeV}) + \alpha \,(\qty{3.5}{\MeV}). \label{eq:fusion}
\end{equation}

The products of this reaction are emitted at \qty{180}{\degree} with respect to each other in the COM frame. Alpha particles are detected with a position-sensitive scintillation detector located close to the target. The associated neutrons travel in the opposite direction of the alpha particles and towards the measurement volume where they interact with the target material. The interaction then produces gamma rays that are detected in nearby gamma-ray detectors. For coincident alpha/gamma-ray pairs, timestamps, energies, and alpha positions are recorded and used to reconstruct the inelastic scattering locations of the neutrons. 

Due to the use of hazardous tritium, API generators are constructed as a sealed source and operated with a mixture of deuterium and tritium gas. Our neutron generator was loaded with equal parts deuterium and tritium. The beam extracted from the ion source consists of atomic and molecular deuterium and tritium ions (D$^+$, T$^+$, D$_2^+$, T$_2^+$, DT$^+$). The actual ratio of atomic to molecular ions cannot be monitored during generator operation and thus is not known. However, atomic fractions in the range of \qtyrange{60}{90}{\%} were reported in the literature for similar microwave-driven ion sources\cite{vainionpaa2013high, ji2009initial}.

In a basic two-particle collision, the COM velocity will be completely defined by the beam direction and beam energy. However, since we have a thick target, the exact COM velocity for each fusion event in API is not known because the incoming ions lose energy in the titanium layer before reacting at different depths and energies. Furthermore, different ion species in the beam will have different initial momenta when entering the titanium layer that will also result in different velocity distributions. Together this leads to a complex distribution of COM velocities for the fusion reactions. The distribution, however, is still completely determined by the initial beam energy and geometry of the setup (assuming other parameters, such as the mixture of ion species, the target material, etc. are fixed). We describe how we simulate this distribution and how the $X$, $Y$, $Z$ position errors resulting from the COM velocity uncertainties are analyzed.

\subsection{Reconstruction Method Including COM Motion}

\begin{figure}[ht]
\centering
\includegraphics[width=\linewidth]{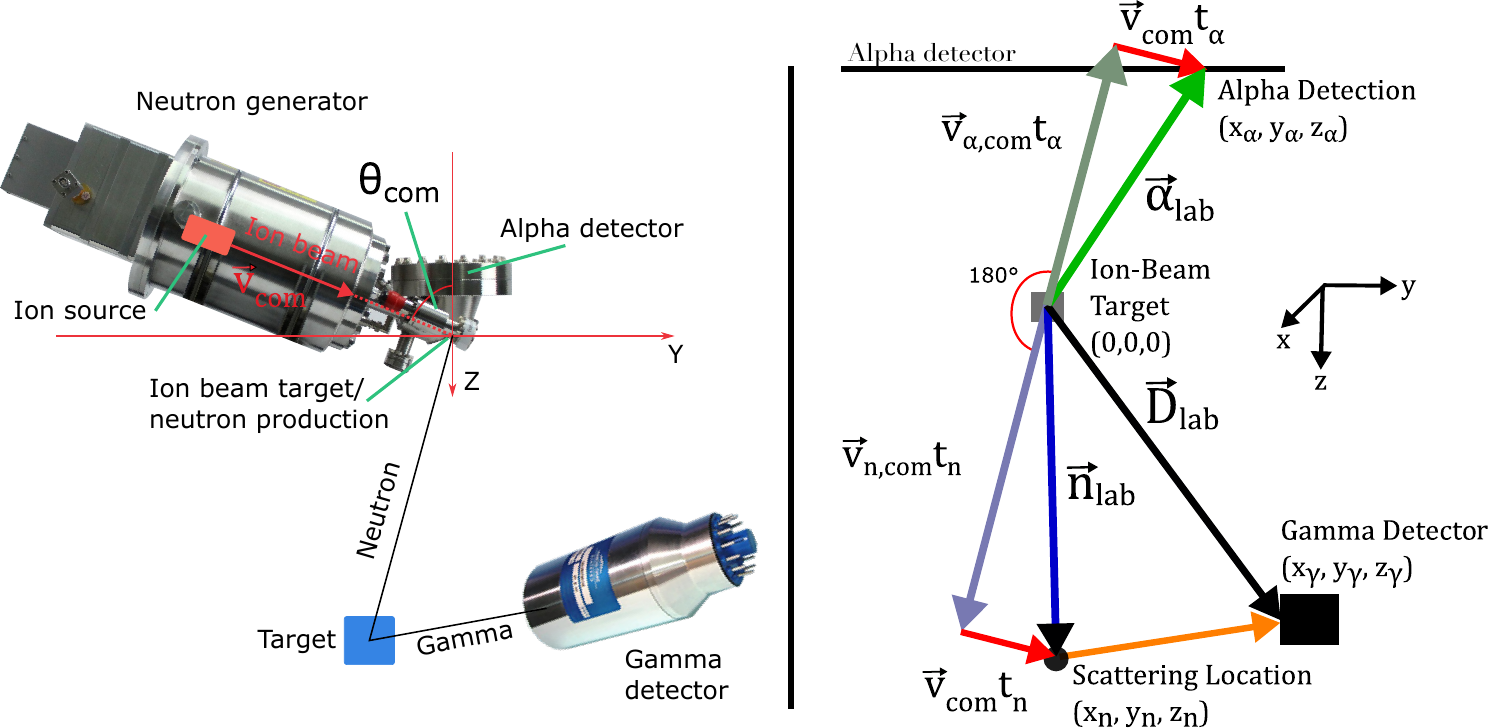}
\caption{Left: schematic of the setup. Showing the neutron generator, the alpha detector (positioned parallel to the ground, i.e., to the $XY$ plane), the ion source, and the ion beam (angled in respect to the ground). Right: Vectors indicating positions of a neutron and alpha particle produced in a DT fusion reaction that then hit the alpha detector and an object where the neutron produces a gamma ray. The schematic visualizes positions and velocities as used in the text. Note that the angle between the alpha vector, $\vec{\alpha}_{lab}$, and neutron vector, $\vec{n}_{lab}$, is less than \qty{180}{\degree} in the lab frame due to the COM velocity, $\vec{v}_{com}$, which in our setup lies in the $YZ$ plane. $t_a$ and $t_n$ are the alpha and neutron travel times, respectively.}
\label{fig:neutron-reconstruction-diagram}
\end{figure}

Reconstructing the location where the neutron scattered requires a measurement of the position where the alpha particle hits the detector and of the time difference between detection of alpha particle and coincident gamma ray. Figure~\ref{fig:neutron-reconstruction-diagram} is a not-to-scale illustration of the reconstruction that is described in the following paragraphs.

To reconstruct the neutron location, we first calculate the alpha travel time, $t_\alpha$, from the alpha-particle position, $\vec{\alpha}_{lab}$, and COM velocity, $\vec{v}_{com}$ 
\begin{equation}
\begin{split}
t_\alpha =&  \frac{- \vec{\alpha}_{lab}\cdot\vec{v}_{com}}{v_{\alpha, com}^2 - v_{com}^2} \label{eq:alpha_reconstruction}\\
&\varpm \frac{\sqrt{(\vec{\alpha}_{lab}\cdot\vec{v}_{com})^2 +  |\vec{\alpha}_{lab}|^2 \left(v_{\alpha, com}^2 - v_{com}^2\right)} }{v_{\alpha, com}^2 - v_{com}^2}
\end{split}
\end{equation}
where $\vec{v}_{\alpha, com}$ is the velocity of the alpha particle in the COM system.
Equation~\eqref{eq:alpha_reconstruction} is the solution to a quadratic equation \cite{Ayllon_Unzueta2020}, in which only the positive branch results in physically meaningful results. Additionally, we note that the above solution assumes $|\vec{v}_{com}| <|\vec{v}_{\alpha, com}|$, which is always true for our API generator.

Next, the alpha travel time, $t_\alpha$, the COM velocity, $\vec{v}_{com}$, and the alpha detection position, $\vec{\alpha}_{lab}$, are used to find the velocity vector of the alpha particle in the COM frame, $\vec{v}_{\alpha,com}$
\begin{equation}
\vec{v}_{\alpha,com} = \frac{\vec{\alpha}_{lab}}{t_\alpha} - \vec{v}_{com} \label{eq:alpha_velocity}.
\end{equation}

\begin{figure*}[tbh!]
\begin{equation}
t_n = \frac{- \Delta{t}\, c^2 + \vec{D}_{lab}\cdot\vec{v}_{n, lab} }{|\vec{v}_{n, lab}|^2 - c^2} \varpm \label{eq:neutron_reconstruction} 
\frac{\sqrt{({\Delta{t}}\, c^2 - \vec{D}_{lab}\cdot\vec{v}_{n, lab})^2 - (|\vec{v}_{n, lab}|^2 - c^2)(|\vec{D}_{lab}|^2 - \Delta{t}^2 c^2)}}{|\vec{v}_{n, lab}|^2 - c^2}
\end{equation}
\hrulefill
\end{figure*}

Because the alpha particle and neutron travel in opposite directions in the COM frame, we can calculate the neutron velocity by inverting the alpha velocity and scaling it by the known velocity $|\vec{v}_{n,com}|$. Then, the COM velocity is added back to find the neutron velocity in the lab frame, $\vec{v}_{n, lab}$
\begin{equation}
\vec{v}_{n, lab} = -|\vec{v}_{n,com}|\frac{\vec{v}_{\alpha,com} }{|\vec{v}_{\alpha,com}|} + \vec{v}_{com} \label{eq:neutron_velocity}.
\end{equation}

The measured time interval between alpha and gamma detection is then used to find the neutron travel time, $t_n$. First, we add the previously calculated alpha travel time, $t_\alpha$, to the measured time interval between alpha and gamma detection, $t_{\text{measured}}$, to get the sum, $\Delta{t}$, of the gamma travel time, $t_\gamma$, and the neutron travel time, $t_n$
\begin{equation}
\Delta{t} = t_{\text{measured}} + t_\alpha = t_n + t_\gamma. \label{eq:flight_time}
\end{equation}

Next, equation~\eqref{eq:neutron_reconstruction} uses the neutron velocity in the laboratory frame, $\vec{v}_{n, lab}$, the position of the gamma detector, $\vec{D}_{lab}$, and the combined time-of-flight of the gamma ray and the neutron, $\Delta{t}$, to find $t_n$.
Similar to equation~\eqref{eq:alpha_reconstruction}, this is a solution to a quadratic equation, and only one solution makes sense physically (the other solution gives negative travel times for the alpha particle, gamma ray, or neutron).

Using the travel time and the neutron velocity vector, equation \eqref{eq:scattering_center} finally gives the location of the scattering center, 
\begin{equation}
\vec{n}_{lab} = t_n \vec{v}_{n, lab}. \label{eq:scattering_center}
\end{equation}

\subsection{Energy Loss and Reaction Probabilities in Thick Targets}

In order to examine how the reconstruction algorithm changes due to the COM movement, we first consider the fusion reactions in the target in more detail. We have already described the five different ions that can lead to a neutron production event; it is important to consider that the ion impinging on the target may be in a molecular state (D$_2^+$, T$_2^+$, or DT$^+$) where the molecule splits apart upon collision, forming two ions each with a fraction of the energy when compared to the atomic ions (D$^+$ or T$^+$). Since no measurement of the distribution of ion species in the extracted beam is available for our generator and we cannot measure it ourselves since the generator is sealed, we assume here that the proportion of each species of molecular ion is equal, i.e., 25\% D$_2^+$, 25\% T$_2^+$, and 50\% DT$^+$. We make this assumption because the generator is filled with equal parts deuterium and tritium and we also assume here the same ionization probabilities for the different isotopes. The reduction in projectile energy due to molecular ions reduces the COM velocity as well as the probability of neutron production. The ion source of our generator produces mostly atomic ions, ranging from 50\% to 90\% depending on operating conditions\cite{ji2009initial, ji2015development}. However, because we cannot monitor the atomic fraction in actual operation, we assume in our calculations below that on average 70\% of the ions are atomic. We show in Section~\ref{sec:reconstruction} how the position (angular) error changes with atomic fraction.

\begin{figure}[ht]
\centering
\includegraphics[width=0.9\linewidth]{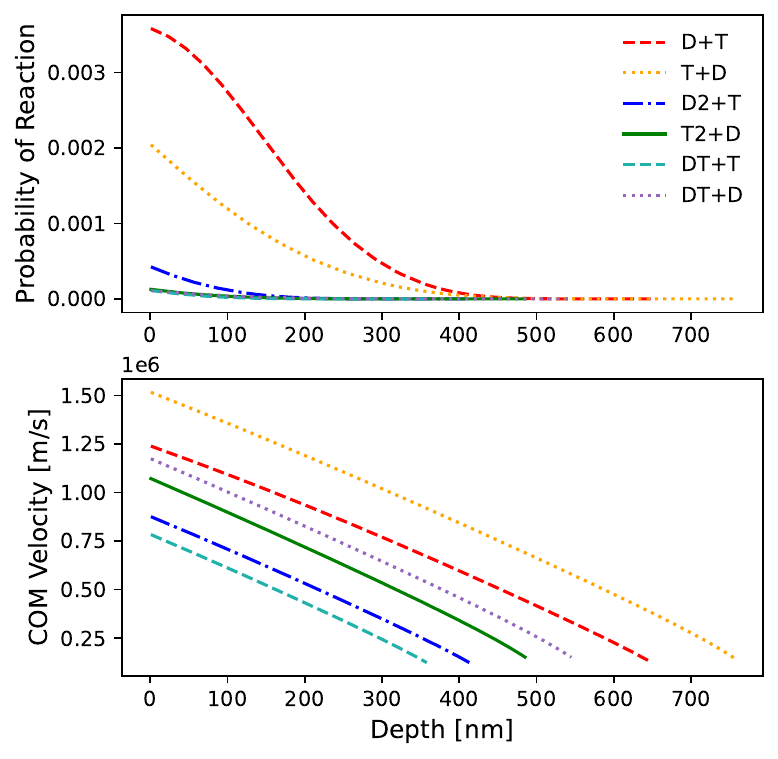}
\caption{Neutron production probability (top) and COM velocity (bottom) plotted as a function of depth for the six different projectile-target combinations (DT$^+$ on T and DT$^+$ on D are considered separately), using an acceleration voltage of \qty{100}{\keV} and assuming 70\% of the ions are atomic and a 50:50 mix of deuterium and tritium. The probability is found by weighting the cross section according to the proportion of atomic to molecular ions. In the listed reaction the second particle is assumed to be at rest in the target material.}
\label{fig:probability-vcm-depth}
\end{figure}

Before weighting the contributions of reactions with different COM velocities, we first examined the fusion reaction probability as a function of the depth the ion travels into the target before it reacts. As the ion travels further into the target material, it loses energy due to nuclear and electronic stopping, causing the fusion cross section and the COM velocity to correspondingly decrease. We modeled the energy loss with the widely-used computer program Stopping and Range of Ions in Matter (SRIM)\cite{srim}. To automate parameter scans in SRIM, we used pysrim\cite{pysrim}, a python interface for SRIM. We created a distribution of the possible ion energies using these energy-loss simulations. Together with the energy-dependent cross section of the fusion reaction we obtained COM velocities and energies for each step that the ion travels into the titanium layer until the cross section became negligible. This was done for all contributing ions, and the resulting distributions are shown in Fig.~\ref{fig:probability-vcm-depth} for an acceleration voltage of \qty{100}{\keV}. As expected, the fusion reactions are dominated by atomic ions, which have higher COM velocities than molecular ions.

\begin{figure}[ht]
\centering
\includegraphics[width=0.9\linewidth]{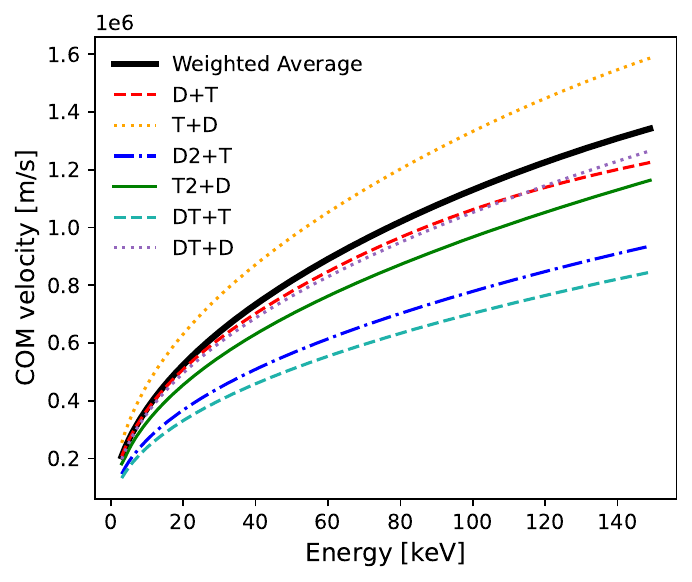}
\caption{Average COM velocity for each projectile-target combination as a function of ion beam energy. The average shown is weighted by the cross section assuming 70\% atomic ions and a 50:50 mixture of deuterium and tritium.}
\label{fig:vcm-energy}
\end{figure}

From the probabilities shown in Fig.~\ref{fig:probability-vcm-depth}, an average COM velocity can be calculated for a given acceleration voltage by weighting the COM velocities for a certain depth with the reaction cross section for that depth. This calculated average COM velocity for each of the potential reactions is shown in Fig.~\ref{fig:vcm-energy} as a function of ion beam energy. The solid black line in Fig.~\ref{fig:vcm-energy} shows the weighted average assuming 70\% atomic ions and a 50:50 mixture of deuterium and tritium that is used later on in the reconstruction.

In this work we investigate two different methods of including the effect of the COM velocity. The first method is to use the weighted average shown in Fig.~\ref{fig:vcm-energy} which gives a single value for the COM velocity at a given  ion beam energy, and therefore the reconstruction can be performed with this average velocity. The second, more detailed method is to use the results shown in Fig.~\ref{fig:probability-vcm-depth} which gives a distribution of probabilities and velocities for a given ion energy. The reconstruction can then be done for the distribution of COM velocities and then be weighted and averaged later to gain a single neutron position and also to calculate a spread in reconstructed positions. The differences between the two reconstruction methods are described in detail later.  

\section{Reconstruction Analysis and error estimation}
\label{sec:reconstruction}

In this section, we conduct the reconstruction in order to investigate the impact of the COM movement. First, the two reconstruction methods were applied to a single event with a given alpha position ($X=\qty{1}{\cm}$, $Y=\qty{1}{\cm}$) and travel time ($t_{\text{measured}} =\qty{15}{\ns}$).
These reconstructions were calculated using the aforementioned assumptions for ion beam compositions, an acceleration voltage of \qty{100}{\kV}, and an angle between beam and alpha detector, otherwise known as the COM angle, of \qty{67.5}{\degree} which is the angle used in our system (see Fig.~\ref{fig:neutron-reconstruction-diagram}).
\begin{figure}[ht]
\centering
\includegraphics[width=1\linewidth]{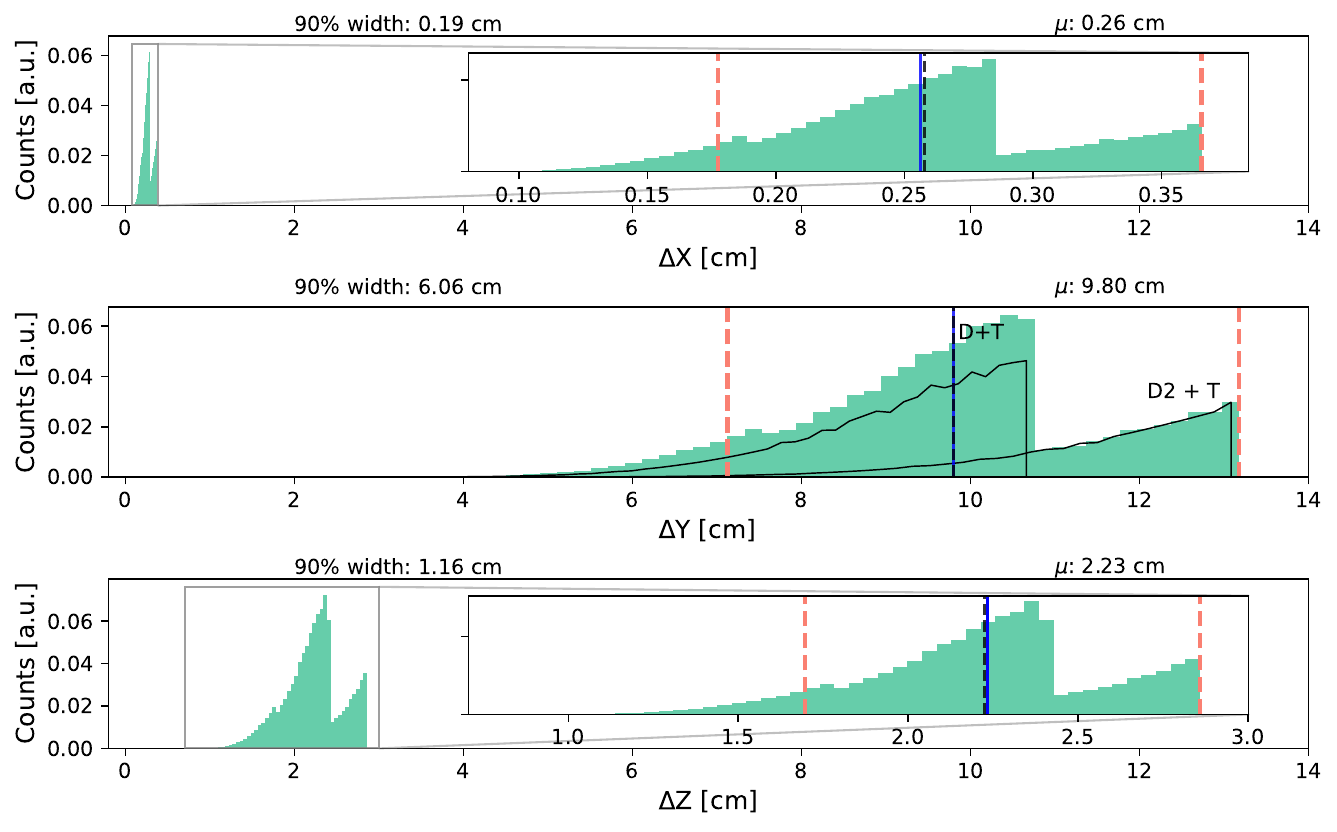}
\caption{The distribution of scattering locations of a neutron scattering event for a given alpha detection position and time interval. For the $\Delta{Y}$ scattering locations, we plot distributions representing the main contributions from the different beam ions (plotted separately in black in the middle plot for the main projectile-target couples). The mean, $\mu$, and width of 90\% of the distribution are given by the black and red dashed lines, respectively, and the solid blue lines represent a reconstruction using the averaged COM velocity from Fig.~\ref{fig:vcm-energy}.}
\label{fig:singlepoint-distribution}
\end{figure}

Figure~\ref{fig:singlepoint-distribution} shows the results of the two reconstruction approaches. A single position is calculated using the average COM velocity (solid blue line) while the second method results in a distribution of $X$, $Y$, $Z$ positions. The average position of the distribution (dashed black line) differs only slightly from the values calculated using the averaged COM velocity. The distributions, seen in Fig.~\ref{fig:singlepoint-distribution}, exhibit peaks relating to the underlying COM velocity distributions for each of the reaction types.  We define the error in the reconstructed position to be the interval covering 90\% of the distribution, taken from the leading edge (marked by dashed red lines in the figure). We use this measure because the distribution has a strong non-Gaussian shape. As a result of this, resolution measurements that quote Gaussian 1-$\sigma$ values cannot directly be compared to this measure. As expected, the largest offset (\qty{9.40}{\cm}) as well as error (\qty{6.12}{\cm}) is in the $Y$-direction since it is the largest component of the COM velocity for our API system. The errors in $Z$ (\qty{1.22}{\cm}) and $X$ (\qty{0.20}{\cm}) are much smaller than the error in $Y$ for the given alpha position and travel time. It should be noted that although the COM velocity has no $X$ component, it still introduces an error in $X$ during the reconstruction. This is due to the COM corrections causing a variation in alpha travel time, which affects all three Cartesian components in the reconstruction.

We note that the $Y$ distribution ($\sim$30$\times$ wider than the $X$- and $\sim$5$\times$ wider than the $Z$-distributions) looks the same for all alpha positions, whereas the distributions in $X$ and $Z$ can exhibit different shapes (not shown, but example plots are available in \cite{zenodo}). For example, although all distributions show multiple-peaks as seen in Fig.~\ref{fig:singlepoint-distribution}, for some alpha positions the $X$ and $Z$ distributions can appear very narrow or flipped. This effect results from the fact that the alpha position directly affects the alpha travel time; changes in the alpha travel time affect the neutron travel time, which then results in a different distribution of scattering centers. 

\begin{figure}[ht]
\centering
\includegraphics[width=1\linewidth]{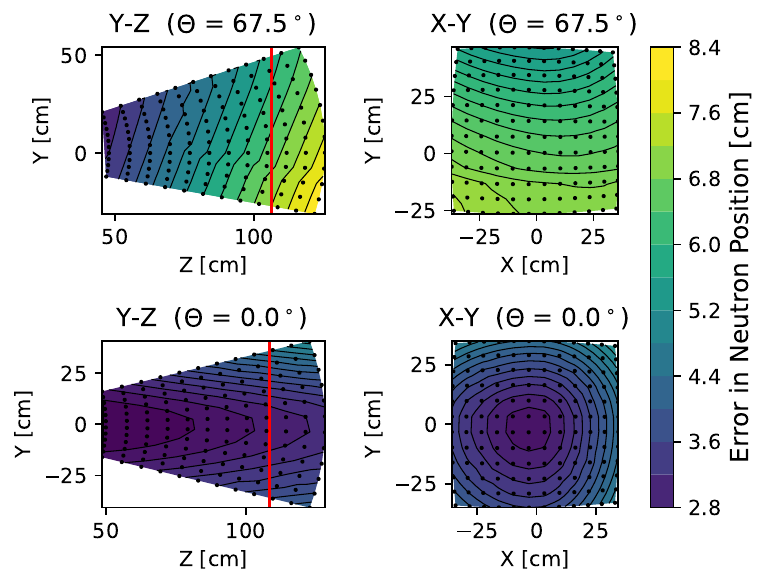}
\caption{A contour plot of the error in the reconstruction within the measurement volume in $YZ$ and $XY$ for COM angles of \qty{67.5}{\degree} and \qty{0}{\degree}. The vertical red line on the $YZ$ plots shows the location of the $XY$ slice.}
\label{fig:comangle}
\end{figure}

In order to capture this effect across the entire volume that we aim to measure with the API system, we created an equally spaced mesh of alpha positions and times between alpha and gamma detection to generate many data points and carry out the same calculations as done for Fig.~\ref{fig:singlepoint-distribution}. For each point, the 90\% interval in the three directions were added in quadrature, then the square root was taken to get a total position error which is plotted in Fig.~\ref{fig:comangle}.

The total error of the reconstructed neutron locations approaches a maximum error of about \qty{6}{\cm} at a depth (defined as the $Z$-distance) of \qty{1.2}{\m} as shown in Fig.~\ref{fig:comangle}. Again, this error is defined as a 90\% interval of the distribution and therefore not directly comparable to the 1-$\sigma$ or FWHM of a Gaussian-like distribution. The error increases in $YZ$ due to the direction of the COM velocity in our setup. The angular resolution of an API system is limited by several factors apart from the COM effect we are discussing here: mainly the size of the beam spot on the target and the alpha detector position resolution. The quoted beam spot size for our generator at full acceleration voltage is \qty{2}{\mm} in diameter. However, at lower operating voltages, it can be significantly larger which results in an angular resolution between \qty{2}{\degree} and \qty{4}{\degree} for our generator\cite{Unzueta_Mauricio2021-jq}. Furthermore, the timing resolution between alpha and gamma-ray detectors, \qty{1.25}{\ns} for our system\cite{Unzueta_Mauricio2021-jq}, corresponds to a depth resolution of \qty{6}{\cm}. This uncertainty also contributes to errors in $X$ and $Y$, since theses values are coupled to the measured time in equation \eqref{eq:neutron_reconstruction}. The error from the distribution of COM velocities is of the same magnitude (but still smaller) compared to the uncertainties from the system's angular and depth resolutions in $Y$, but in $X$ and $Z$ the COM contributions are much smaller. We also note that the resolution measurements \cite{Unzueta_Mauricio2021-jq} were taken with an earlier version of the Adelphi API neutron generator with slightly different geometry. We expect the newer version to have a better resolution (for example, due to less neutron target material and therefore less scattering in the target), but have not quantified this in detail. 

The direction of the COM velocity has a significant impact on the magnitude of errors in the neutron scatter location reconstruction. These errors can be improved by reducing the COM angle. Figure~\ref{fig:comangle} shows that at \qty{0}{\degree} the errors are greatly reduced compared to our current setup at \qty{67.5}{\degree}. However, achieving smaller angles is not straight forward, since the alpha detector cannot be allowed to block the ion beam. Minimizing the error may be possible by placing the alpha detector at \qty{0}{\degree} (orthogonal to the ion beam) and using a small hole in the detector for the ion beam to pass through. A small angle has the additional benefit of decreasing alpha scattering in the neutron production target.

\begin{figure}[ht]
\centering
\includegraphics[width=0.9\linewidth]{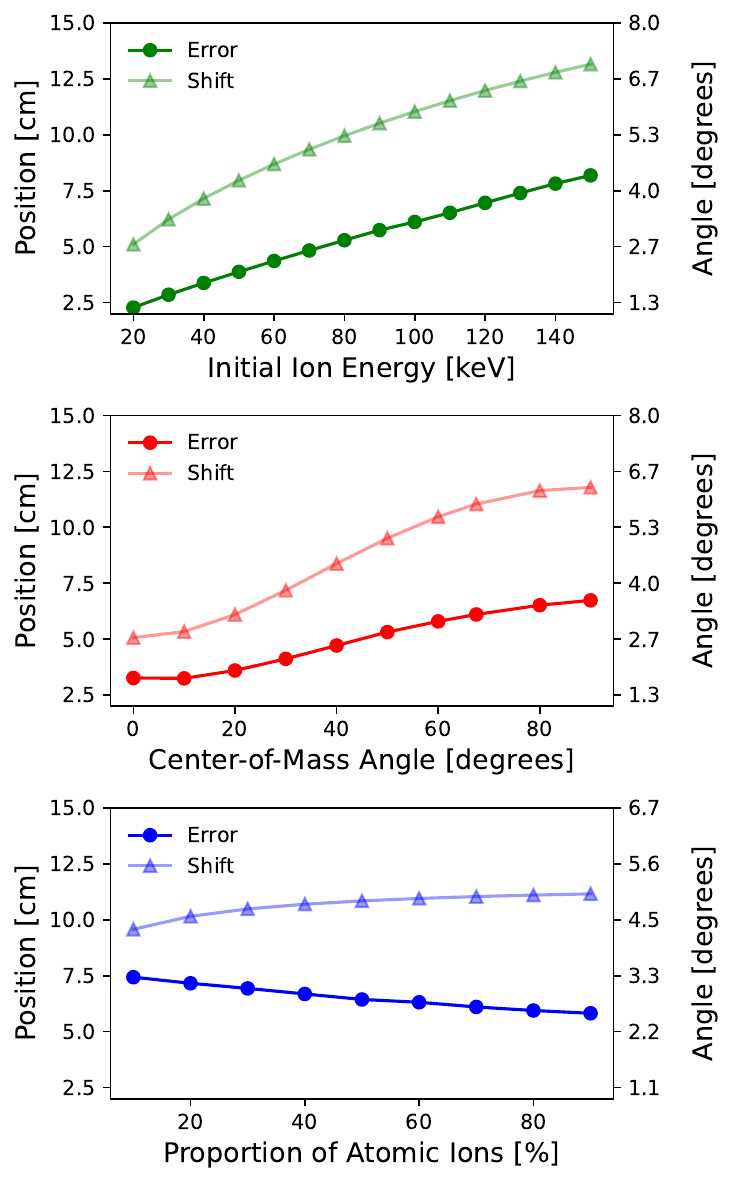}
\caption{Averaged reconstruction error and shift/tilt at a depth $Z$ = \qty{1.0}{\meter} from the neutron source as a function of ion energy (top), COM angle  (middle), and atomic percent (bottom). In each case, the other two parameters are fixed, using fixed values of 70\% atomic percentage, \qty{100}{\keV} and \qty{67.5}{\degree}, accordingly.}
\label{fig:energies-percents-plots}
\end{figure}

We additionally explored the effect of varying multiple parameters which we kept fixed in our previous simulations, such as changing the acceleration voltage between \qty{20}{\kV} and \qty{150}{\kV}, changing the COM angle between \qty{0}{\degree} to \qty{90}{\degree}, and changing the proportion of atomic ions from 10\% to 90\%. To compare the results, we averaged the total error in $X$ and $Y$ (assuming a \qtyproduct{5x5}{\cm} detector) and plotted the error at $Z=\qty{100}{\cm}$ for the varying parameter values, as well as the resulting shift/tilt. As seen in Fig.~\ref{fig:energies-percents-plots} (top), the error and shift/tilt increased with the energy.  In Fig.~\ref{fig:energies-percents-plots} (middle), the effect of the COM angle can clearly be seen such that the error was reduced by decreasing this angle. In Fig.~\ref{fig:energies-percents-plots} (bottom), the error in the reconstruction depended only weakly on the ratio between atomic and molecular ions in the beam. For this reason, the results presented in this article for the specific value of 70\% will be also valid for an ion beam  with an atomic fraction above 50\% and therefore most API systems that use a microwave-driven ion source.

\section{Experimental Data}
We attempted to measure the effect of COM on our reconstruction using a thin carbon (graphite) slab (\qtyproduct{1x10x20}{\cm}) aligned with the short width in the direction of interest (i.e., the \qtyproduct{1x20}{\cm} face was in the $XY$ plane and the \qty{1}{\cm} edge aligned in $X$- or $Y$-direction) . The slab was thin enough compared to our resolution to give a good estimation of resolution of our API system. We aligned the center of the slab directly below the neutron source at $X=0$  and $Y=0$ and positioned it \qty{51}{\cm} below the neutron source. The carbon slab was  placed on low-density foam material to separate it from the high-density floor. 

\begin{figure}[ht]
\centering
\includegraphics[width=1\linewidth]{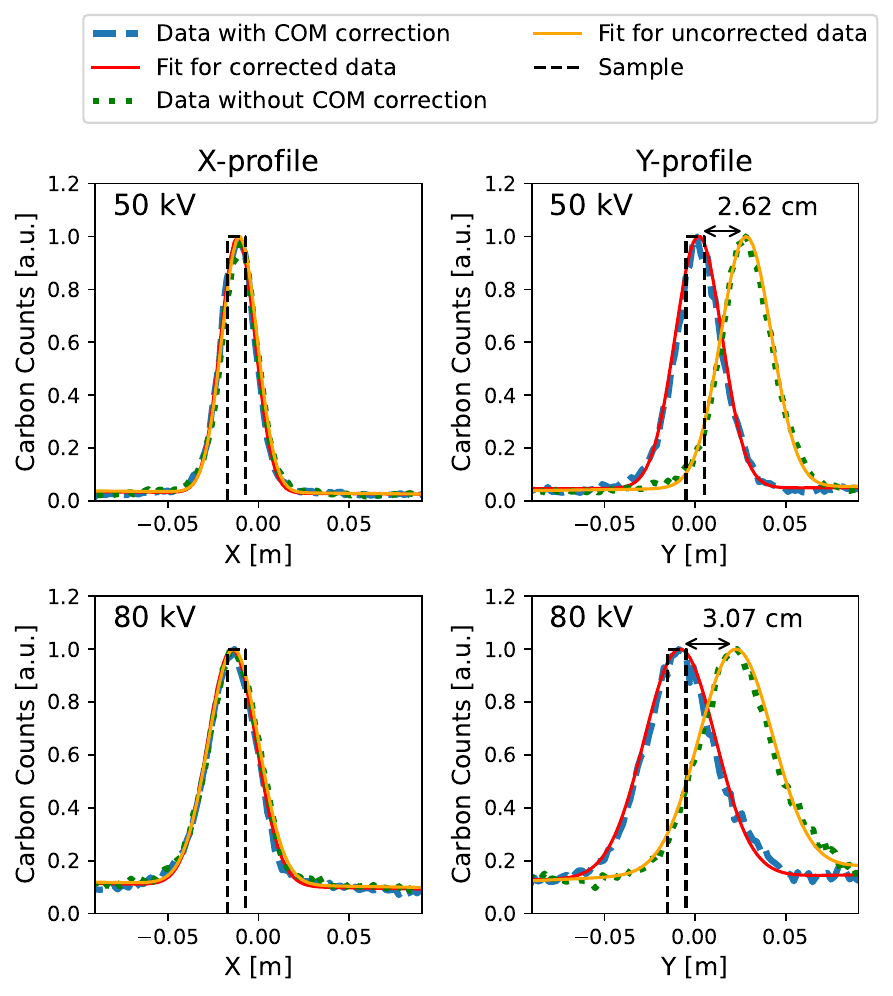}
\caption{Histograms of carbon-slab data taken at different voltages and orientations. As expected, the plots show that the resolution is worse in the $Y$-direction than in the $X$-direction and worse at \qty{80}{\keV} than at \qty{50}{\keV}. The FWHM of the $X$-profile was \qty{2.2}{\cm} and \qty{3.1}{\cm} at \qty{50}{\keV} and \qty{80}{\keV}, respectively. Meanwhile, the $Y$-profile had a FWHM of \qty{3.0}{\cm} and \qty{4.6}{\cm} at \qty{50}{\keV} and \qty{80}{\keV}. We also plot the data without the COM correction, and it showed a visible shift in $Y$ but not $X$. The carbon slab location is indicated by the dashed black line.}
\label{fig:data}
\end{figure}

Data was taken at acceleration energies of \qty{50}{\keV} and \qty{80}{\keV} with run times of 5 hours and 2 hours respectively.  We then reconstructed the $XYZ$-positions both with and without COM corrections. For the data analysis, only counts inside a region of interest around the carbon-slab (selected by $X$, $Y$, and $Z$ ranges) and an energy interval around the carbon peak of \qtyrange{3}{5}{\MeV} were tallied. From Fig.~\ref{fig:data}, we can see that the COM correction causes a shift in $Y$ but not in $X$, which conforms to the fact that our ion beam moves in the $YZ$ plane. Although the center of the slab is placed at $X=0$ and $Y=0$ for each respective profile, we measure a reconstructed center position that is slightly off (roughly \qtyrange{1}{2}{\cm}). We attribute this to other effects that influence our reconstruction, such as the beam not being centered on the target, the alpha detector not being level with the floor, etc. Overall the reconstructed position agrees with the $XY$ origin within our measurement errors. As expected, we see that ignoring the COM-based reconstructions leads to a shift in the $Y$ data of several centimeters.
However, we expected to see a \qty{1.6}{\cm} larger shift for \qty{80}{\keV} than for \qty{50}{\keV} in $Y$  that we do not see in our measurement. We attribute this to a change in the beam position on target as a function of the acceleration voltage (which is expected if the beam is not directly on axis) that could counteract the COM shift.  We also see a broader peak in $Y$ at \qty{80}{\keV} than at \qty{50}{\keV} with FWHM values of \qty{4.6}{\cm} and \qty{3.0}{\cm}, respectively. However, the increase is larger than we expect from our COM calculations.  Most likely the observed change in FWHM is dominated by a varying ion beam spot size (which is not surprising, since beam focusing and thus beam spot size likely vary with acceleration voltages).

\section{Conclusion}
We have described COM corrections in an API setup and analyzed the dependencies of the corrections and their associated errors on ion beam energy and atomic fraction, and on the angle between ion beam and alpha detector. 

We compared using an average COM velocity in the scattering position reconstruction to first calculating the full distribution and then taking its mean. As can be seen in Fig.~\ref{fig:singlepoint-distribution}, using the average COM velocity gives almost the same position as the mean of the full distribution. Over our volume of interest (in our case \qty{50}{\cm} to \qty{120}{\cm} in $Z$ and roughly $\qty{50}{\cm}\times\qty{50}{\cm}$ in $X$ and $Y$ at $Z=\qty{50}{\cm}$), the largest difference between the two reconstruction methods is less than \qty{0.15}{\mm} \cite{zenodo} and is therefore insignificant (compared to a resolution of several centimeters). Since using the average COM velocity and performing only a single reconstruction is computationally much faster than calculating the full distribution, the former method ought to be preferred for analyzing experimental data.

Using the averaged COM velocity to reconstruct scattering locations can correct for a shift/tilt of the image as much as 12 centimeters at a distance of around \qty{1}{\meter} from the neutron source (see Fig.~\ref{fig:energies-percents-plots}, top plot at \qty{100}{\keV}). Event-by-event calculations, in contrast, allowed us to quantify the impact of the distribution of COM velocities on the spatial resolution. The additional uncertainty in the FWHM arises due to multiple factors -- the presence of atomic and molecular ions in the beam, whether a deuterium ion impinges on a tritium ion or vice versa, and the variation of the ion's energy due to the energy loss within the target.

In our current setup, the COM error in the reconstructed position is largest in the beam direction ($Y$ axis) as seen in Fig.~\ref{fig:comangle} (top left). The $Y$-position error (defined as a region spanning 90\% of the distribution) is about \qty{3.4}{\centi\meter} at a distance of \qty{60}{\centi\meter} from the neutron source \cite{zenodo}. In the $X$- and $Z$-directions, the COM correction error is an order of magnitude less than in $Y$ (and the error in $X$ is normally much smaller than in $Z$). For our existing API system, the measured position resolution (FWHM) is about \qty{4}{\centi\meter} in $X$ and $Y$ and about \qty{6}{\centi\meter} in $Z$ at a distance of \qty{60}{\cm} \cite{Unzueta_Mauricio2021-jq}. Therefore, the errors in $X$ and $Z$ introduced by the COM movement uncertainty are small compared to the overall error and can be ignored for our setup and data analysis. 
In the beam direction ($Y$) the error is the largest, but in practice our measurements have shown that for our system it is only a small contribution to our measured resolution. Other effects, most likely beam spot size, have a greater effect on the resolution. Future improvements in alpha-detector and timing resolution as well as ion beam spot size on the target may make the error due to the COM velocity spread more relevant. This error can be reduced (see Fig.~\ref{fig:energies-percents-plots}) by decreasing the COM angle, i.e., the angle between alpha detector and ion beam, and future API neutron generator designs might need to incorporate a smaller COM angle to achieve a higher resolution.
In summary, for current API systems, the COM correction should be included to correct for a shift/tilt, but contributions to the resolutions can currently still be ignored. To correct for the COM shift/tilt the averaged COM velocity can be used and a detailed calculation is not necessary.

\section*{Acknowledgments}
The information, data, or work presented herein was funded by the Department of Energy (DOE) through Lawrence Berkeley National Laboratory's Laboratory-Directed Research and Development (LDRD) grant as part of its Carbon Negative Initiative under Contract No. DE-AC02-05CH11231.

\bibliographystyle{IEEEtranDOI}
\bibliography{paper}

\begin{thebibliography}{10}
\providecommand{\url}[1]{#1}
\csname url@samestyle\endcsname
\providecommand{\newblock}{\relax}
\providecommand{\bibinfo}[2]{#2}
\providecommand{\BIBentrySTDinterwordspacing}{\spaceskip=0pt\relax}
\providecommand{\BIBentryALTinterwordstretchfactor}{4}
\providecommand{\BIBentryALTinterwordspacing}{\spaceskip=\fontdimen2\font plus
\BIBentryALTinterwordstretchfactor\fontdimen3\font minus
  \fontdimen4\font\relax}
\providecommand{\BIBforeignlanguage}[2]{{%
\expandafter\ifx\csname l@#1\endcsname\relax
\typeout{** WARNING: IEEEtran.bst: No hyphenation pattern has been}%
\typeout{** loaded for the language `#1'. Using the pattern for}%
\typeout{** the default language instead.}%
\else
\language=\csname l@#1\endcsname
\fi
#2}}
\providecommand{\BIBdecl}{\relax}
\BIBdecl

\bibitem{vainionpaa2013high}
\BIBentryALTinterwordspacing
J.~H. Vainionpaa, J.~L. Harris, M.~A. Piestrup, C.~K. Gary, D.~L. Williams,
  M.~D. Apodaca, J.~T. Cremer, Q.~Ji, B.~A. Ludewigt, and G.~Jones, ``High
  yield neutron generators using the dd reaction,'' \emph{AIP Conference
  Proceedings}, vol. 1525, no.~1, pp. 118--122, 2013. [Online]. Available:
  \url{http://dx.doi.org/10.1063/1.4802303}
\BIBentrySTDinterwordspacing

\bibitem{carasco2008field}
\BIBentryALTinterwordspacing
C.~Carasco, B.~Perot, S.~Bernard, A.~Mariani, J.-L. Szabo, G.~Sannie, T.~Roll,
  V.~Valkovic, D.~Sudac, G.~Viesti \emph{et~al.}, ``In-field tests of the
  euritrack tagged neutron inspection system,'' \emph{Nuclear Instruments and
  Methods in Physics Research Section A: Accelerators, Spectrometers, Detectors
  and Associated Equipment}, vol. 588, no.~3, pp. 397--405, 2008. [Online].
  Available: \url{"https://dx.doi.org/10.1016/j.nima.2008.01.097"}
\BIBentrySTDinterwordspacing

\bibitem{fontana2017detection}
\BIBentryALTinterwordspacing
C.~L. Fontana, A.~Carnera, M.~Lunardon, F.~Pino, C.~Sada, F.~Soramel,
  L.~Stevanato, G.~Nebbia, C.~Carasco, B.~Perot \emph{et~al.}, ``Detection
  system of the first rapidly relocatable tagged neutron inspection system
  (rrtnis), developed in the framework of the european h2020 c-bord project,''
  \emph{Physics Procedia}, vol.~90, pp. 279--284, 2017. [Online]. Available:
  \url{"https://doi.org/10.1016/j.phpro.2017.09.010"}
\BIBentrySTDinterwordspacing

\bibitem{HAUSLADEN2007387}
\BIBentryALTinterwordspacing
P.~Hausladen, P.~Bingham, J.~Neal, J.~Mullens, and J.~Mihalczo, ``Portable
  fast-neutron radiography with the nuclear materials identification system for
  fissile material transfers,'' \emph{Nuclear Instruments and Methods in
  Physics Research Section B: Beam Interactions with Materials and Atoms}, vol.
  261, no.~1, pp. 387--390, 2007, the Application of Accelerators in Research
  and Industry. [Online]. Available:
  \url{https://www.sciencedirect.com/science/article/pii/S0168583X07009147}
\BIBentrySTDinterwordspacing

\bibitem{alexakhin2015detection}
\BIBentryALTinterwordspacing
V.~Y. Alexakhin, V.~Bystritsky, N.~Zamyatin, E.~Zubarev, A.~Krasnoperov,
  V.~Rapatsky, Y.~N. Rogov, A.~Sadovsky, A.~Salamatin, R.~Salmin \emph{et~al.},
  ``Detection of diamonds in kimberlite by the tagged neutron method,''
  \emph{Nuclear Instruments and Methods in Physics Research Section A:
  Accelerators, Spectrometers, Detectors and Associated Equipment}, vol. 785,
  pp. 9--13, 2015. [Online]. Available:
  \url{"https://doi.org/10.1016/j.nima.2015.02.049"}
\BIBentrySTDinterwordspacing

\bibitem{kavetskiy2019application}
\BIBentryALTinterwordspacing
A.~Kavetskiy, G.~Yakubova, S.~A. Prior, and H.~A. Torbert, ``Application of
  associated particle neutron techniques for soil carbon analysis,'' in
  \emph{AIP Conference Proceedings}, vol. 2160, no.~1.\hskip 1em plus 0.5em
  minus 0.4em\relax AIP Publishing LLC, 2019. doi: 10.1063/1.5127698 p. 050006.
  [Online]. Available: \url{"https://doi.org/10.1063/1.5127698"}
\BIBentrySTDinterwordspacing

\bibitem{Unzueta_Mauricio2021-jq}
\BIBentryALTinterwordspacing
A.~Unzueta~Mauricio, B.~Ludewigt, B.~Mak, T.~Tak, and A.~Persaud, ``An
  all-digital associated particle imaging system for the {3D} determination of
  isotopic distributions,'' \emph{Rev. Sci. Instrum.}, vol.~92, no.~6, p.
  063305, Jun. 2021. [Online]. Available:
  \url{/url{"https://doi.org/10.1063/5.0030499"}}
\BIBentrySTDinterwordspacing

\bibitem{Sanderman2017-rt}
\BIBentryALTinterwordspacing
J.~Sanderman, T.~Hengl, and G.~J. Fiske, ``\BIBforeignlanguage{en}{Soil carbon
  debt of 12,000 years of human land use},''
  \emph{\BIBforeignlanguage{en}{Proc. Natl. Acad. Sci. U. S. A.}}, vol. 114,
  no.~36, pp. 9575--9580, Sep. 2017. [Online]. Available:
  \url{http://dx.doi.org/10.1073/pnas.1706103114}
\BIBentrySTDinterwordspacing

\bibitem{grogan2010development}
\BIBentryALTinterwordspacing
B.~R. Grogan, ``The development of a parameterized scatter removal algorithm
  for nuclear materials identification system imaging,'' Ph.D. dissertation,
  The University of Tennessee, 2010. [Online]. Available:
  \url{https://trace.tennessee.edu/utk_graddiss/692/}
\BIBentrySTDinterwordspacing

\bibitem{cates2013investigation}
\BIBentryALTinterwordspacing
J.~W. Cates, ``Investigation of time and position resolved alpha transducers
  for {Multi-Modal} imaging with a {D-T} neutron generator,'' Ph.D.
  dissertation, University of Tennessee, Knoxville University of Tennessee,
  Knoxville, 2013. [Online]. Available:
  \url{https://trace.tennessee.edu/utk_graddiss/2406/}
\BIBentrySTDinterwordspacing

\bibitem{adelphi}
``Adelphi technology inc.'' \url{http://adelphitech.com/}, Dec 2021.

\bibitem{ji2009initial}
\BIBentryALTinterwordspacing
Q.~Ji, Y.~Wu, M.~Regis, and J.~W. Kwan, ``Initial testing of a compact portable
  microwave-driven neutron generator,'' \emph{IEEE Transactions on Nuclear
  Science}, vol.~56, no.~3, pp. 1312--1315, 2009. [Online]. Available:
  \url{http://dx.doi.org/10.1109/TNS.2009.2015305}
\BIBentrySTDinterwordspacing

\bibitem{Ayllon_Unzueta2020}
\BIBentryALTinterwordspacing
M.~Ayllon~Unzueta, ``An associated particle imaging system for the
  determination of {3D} isotopic distributions,'' Ph.D. dissertation, UC
  Berkeley, 2020. [Online]. Available:
  \url{https://escholarship.org/uc/item/1h55025b}
\BIBentrySTDinterwordspacing

\bibitem{ji2015development}
\BIBentryALTinterwordspacing
Q.~Ji, B.~Ludewigt, J.~Wallig, W.~Waldron, and J.~Tinsley, ``Development of a
  time-tagged neutron source for snm detection,'' \emph{Physics Procedia},
  vol.~66, pp. 105--110, 2015. [Online]. Available:
  \url{https://dx.doi.org/10.1016/j.phpro.2015.05.015}
\BIBentrySTDinterwordspacing

\bibitem{srim}
\BIBentryALTinterwordspacing
J.~F. Ziegler, M.~D. Ziegler, and J.~P. Biersack, ``Srim--the stopping and
  range of ions in matter (2010),'' \emph{Nuclear Instruments and Methods in
  Physics Research Section B: Beam Interactions with Materials and Atoms}, vol.
  268, no. 11-12, pp. 1818--1823, 2010. [Online]. Available:
  \url{"https://doi.org/10.1016/j.nimb.2010.02.091"}
\BIBentrySTDinterwordspacing

\bibitem{pysrim}
C.~Ostrouchov, ``Python software foundation,''
  \url{https://gitlab.com/costrouc/pysrim}, Jan 2022.

\bibitem{zenodo}
\BIBentryALTinterwordspacing
C.~Egan, A.~Amsellem, D.~Klyde, B.~Ludewigt, and A.~Persaud, ``{Simulation and
  data analysis for "Center-of-Mass Corrections in Associated Particle
  Imaging"},'' June 2023. [Online]. Available:
  \url{https://doi.org/10.5281/zenodo.8302341}
\BIBentrySTDinterwordspacing

\end{thebibliography}

\end{document}